# NeedForHeat DataGear: An Open Monitoring System to Accelerate the Residential Heating Transition

Henri ter Hofte
*Research Group Energy Transition*
*Windesheim University of Applied Sciences*
Zwolle, the Netherlands
henri@terhofte.net

Nick van Ravenzwaaij
*Research Group Energy Transition*
*Windesheim University of Applied Sciences*
Apeldoorn, the Netherlands
nickvanravenzwaaij@protonmail.com

***Abstract*— We introduce NeedForHeat DataGear, an open hardware and open software system for high-resolution residential energy monitoring, designed to accelerate the heating transition before retrofitting. DataGear enables systematic collection of time series data on thermal behaviour of buildings, installations and occupant comfort demand in homes, before heating system upgrades. Its architecture combines modularity, security, low cost, and adaptability for diverse field deployments. We demonstrate its capabilities and utility through three field deployments, showing how pre-transition data can be collected to inform evidence-based retrofit strategies. Unlike conventional solutions centered on home automation or post-installation monitoring, NeedForHeat DataGear emphasizes pre-transition performance insights, equipping researchers and energy professionals with actionable data to support decision-making in the residential heating transition.**

## I. Introduction

The severity of climate change hinges on our cumulative greenhouse gas emissions from this point forward [1]. To avert dangerous consequences, urgent action is needed to achieve net-zero emissions. In the Netherlands, 85% of all homes [2] still relied on setting fire to natural gas for heating in 2024, corresponding to 7.0 of the 8.2 million homes nationwide [3]. Achieving net-zero heating by 2050 therefore requires an average transition rate of more than 22,000 homes per month, sustained over 25 years.

Recent statistics show that in both 2022 and 2023, approximately 23,500 homes per year discontinued the use of natural gas for heating [2]. In other words, the Netherlands must sustain, *each month* for over 25 years, a transition rate comparable to what was recently achieved only *annually*. This stark mismatch underscores the urgency of scaling up efforts and adopting innovative approaches to accelerate the residential heating energy transition.

While cursory observations may suggest opportunities for clustering homes based on building typology, location, or neighbourhood characteristics, a closer examination reveals the dynamic and diverse nature of individual homes. Homes undergo changes over their lifetimes, with varying heating systems, resident preferences, and occupancy patterns. The illusion of uniformity quickly dissipates, underscoring the added value of a tailored, home-specific approach to address the unique challenges posed by the residential heating energy transition [4], [5]. The added value of detailed, home-specific insights is also acknowledged within the contingent approach [6].

Residents facing the heating transition often grapple with key questions, like: *Can I switch to a (hybrid) heat pump without upgrading insulation or replacing my radiators? Do I need to switch to floor heating? What happens if I don't? Will my home remain comfortable, and how will this affect my energy costs? Is the new heating system more sustainable?*

The answers depend on the thermal characteristics of their home and heat distribution system (e.g., radiators), the characteristics of their current and prospective heat generation system, and how well these elements work together to deliver the desired level of comfort. Without tailored insights, residents may delay action, concerned about high costs, insufficient heating, or unclear environmental benefits. Providing home-specific guidance may help accelerate adoption and ensure effective, comfortable transitions.

To illustrate the scale of the potential effect, consider a scenario where 50% of the 7.0 million Dutch homes with gas-fired boilers switched to all-electric heat pumps five years earlier. This could save over 500 PJ final energy, cut emissions by nearly 27 $MtCO_2eq$ (equivalent to removing ~0.5 million passenger cars from the road for five years) and lower household energy costs by more than €6 billion at 2024 prices.

The widespread use of internet-connected energy devices enables data collection for personalized residential heating transition advice. In the Netherlands, smart meter deployment began in 2012, reaching over 87% of small consumer connections by 2021, primarily in individual homes [7]. By 2024, 99% of those aged 12+ had internet access, 95% owned a smartphone, and 31% of homes had smart thermostats [8].

We developed NeedForHeat to leverage this trend and accelerate the residential heating transition, initially for researchers and ultimately for energy advisors. The DataGear subsystem collects data for NeedForHeat Diagnosis [9], which applies physics-informed machine learning to learn household-specific parameters, such as building envelope characteristics, heat distribution system characteristics, and comfort preferences. These insights help identify suitable sustainable heating solutions and assess what-if scenarios for a specific home, such as the impact of insulation upgrades or heating system changes.

By automating data collection and analysis, NeedForHeat streamlines surveying and screening, allowing energy advisors more time to focus on understanding concerns, addressing barriers, and providing tailored guidance to residents. It serves as a comprehensive tool for researchers and advisors, supporting field data collection and diagnostics, which helps them to guide residents towards effective steps in the residential heating transition.

This paper focuses on NeedForHeat DataGear, which gathers data from homes, heating systems, and comfort preferences that we hypothesized to be essential for pre-transition diagnostics. It does not cover NeedForHeat Diagnosis, which uses physics-informed machine learning and heat balance models. In this paper, we ask: *What are the minimal viable features for a low-cost, scalable, privacy-aware system to collect residential data on thermal and energy parameters to inform the home heating transition?*

## II. NeedForHeat DataGear System Requirements

Inspired by studies demonstrating that key building parameters can be learned from monitoring data [10], [11], [12], we launched a residential monitoring study in early 2020 to collect time series data, on indoor temperatures, thermostat setpoints, flow and return temperatures, and gas and electricity use. COVID-19 restrictions emphasized the need for self-installable measurement devices. We evaluated existing commercial and DIY solutions but found limitations:

- Some smart thermostats ([13], [14], [15]) can provide data via APIs but were installed in too few homes.
- A commercially available OpenTherm monitor [16] required a costly gateway, complicating installation.
- DIY alternatives [17], [18] lacked usability and reliability for large-scale deployment.

As no existing solution met our requirements, we developed custom measurement devices for both OpenTherm and other heating systems, without relying on smart thermostats. Integrating with Home Assistant [19] was considered, but dismissed due to its focus on home automation enthusiasts and the added cost and installation complexity of a gateway. Instead, we adopted an open hardware and software approach, prioritizing modularity, extensibility, and ease of use, supported by an app for secure installation.

The next sections outline the system's key requirements.

### A. Core Functional Requirements: the system must

- collect heating-related time series data from pre-transition residential dwellings with gas-fired boilers;
- not interfere with heating control: if the system fails or is removed, heating should work as before;
- support data acquisition from measurement devices and online sources (e.g. connected thermostats);
- support energy transition researchers to collect data, with potential for future use by energy advisors.
- be open-source and open-hardware, with licenses suitable for commercial reuse.

### B. Usability & Deployability: the system must

- be modular, extensible, and configurable, allowing use in different measurement campaigns;
- work in Wi-Fi-connected homes, without requiring an additional gateway or pre-configuration, enabling temporary installation and reuse in another home;
- support easy provisioning, simplifying setup and modification of hardware and software.
- be low-cost: the total cost of measurement devices per home must be under € 250, ideally under € 100.
- be small & lightweight: devices should ideally fit in a PostNL letterbox parcel (max 38x26.5x3.2 cm, 2 kg).
- be easily installable and connectable by residents, including non-technical users.
- be independent from mains power where possible, e.g. by supporting battery-powered operation.

### C. Data Integrity & Privacy Requirements: the system must

- apply privacy-by-design principles:
  - collect only necessary data for research, prioritizing local processing when possible;
  - store only pseudonymized data on the server, no direct identifiers like address, email, or name;
  - restrict pseudonym mapping access to key personnel, to minimize data leakage risks.
- support reliable data acquisition, ensuring:
  - clear distinction between missing & zero values;
  - explicit units for all data points;
  - timestamps for UTC & local time reconstruction;
  - $\geqslant$24 h buffering to handle brief internet outages.

## III. NeedForHeat DataGear Architecture

NeedForHeat DataGear consists of Measurement Devices, Cloud Feeds, the GearUp smartphone app and server software. Fig. 1. provides a high-level overview of the NeedForHeat DataGear architecture, illustrating the data flows from measurement devices like living room module and smart meter module, Cloud Feeds (e.g., Enelogic), the GearUp app, batch import data (e.g., KNMI, Remeha), to the NeedForHeat Server. Measurement devices collect real-world data and transmit it via the existing home Wi-Fi network to the central NeedForHeat Server. Cloud feeds provide real-time or near-real-time data, while batch import data, such as KNMI weather data or advisor-provided diagnostic data, is added later to the server database by campaign advisors, outside the scope of GearUp. All data is processed, analysed, and made available for diagnosis for home heating research or advice.

With the NeedForHeat DataGear open hardware and open software, we support three distinct user groups, each of which will be described in one of the subsequent sections:

- *Residents* of homes participating in research campaigns or seeking energy advice. They install and connect measurement devices and authorize data retrieval from Cloud Feeds via the GearUp app (see section IV).
- *Advisors*, with two sub-roles (not visible in Fig. 1.):
  - *Deployers:* Researchers and energy professionals who configure, deploy and monitor data collection campaigns and devices (see section V).
  - *Developers:* Engineers supporting researchers and energy professionals in extending or modifying the system by developing new or modified hardware or firmware for measurement devices, app or server processes for data retrieval from online services (see section VI).

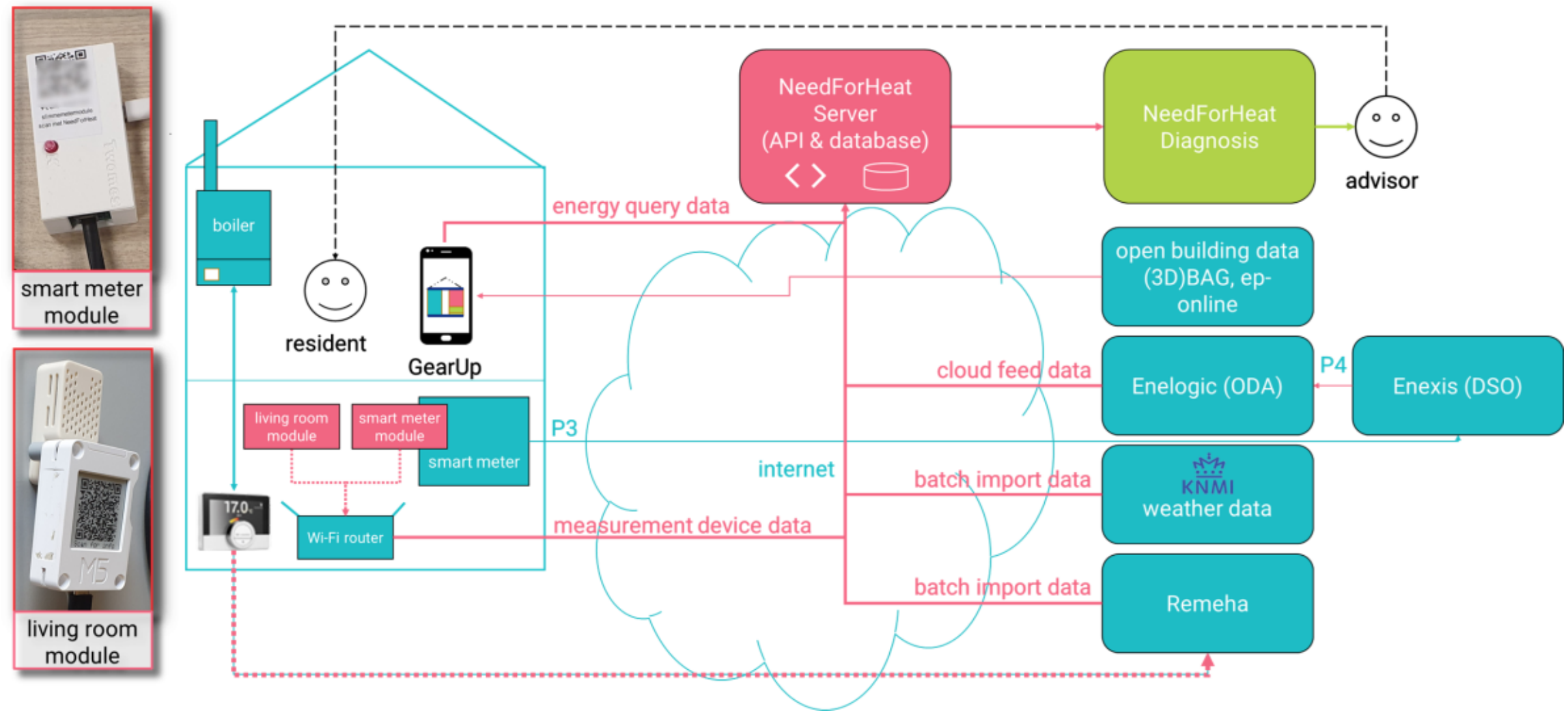

*Fig. 1. NeedForHeat DataGear Architecture*

## IV. Support for Residential Users

NeedForHeat DataGear is designed to enable residents to allow data collection from their home with minimal effort and technical expertise. A key design principle has been maximizing successful self-installation via a smartphone app, reducing the need for professional assistance.

### A. NeedForHeat GearUp App

The NeedForHeat GearUp app is a mobile application for Android and iOS that helps residents install and connect measurement devices, authorize data collection from Cloud Feeds and answer one-time Energy Queries.

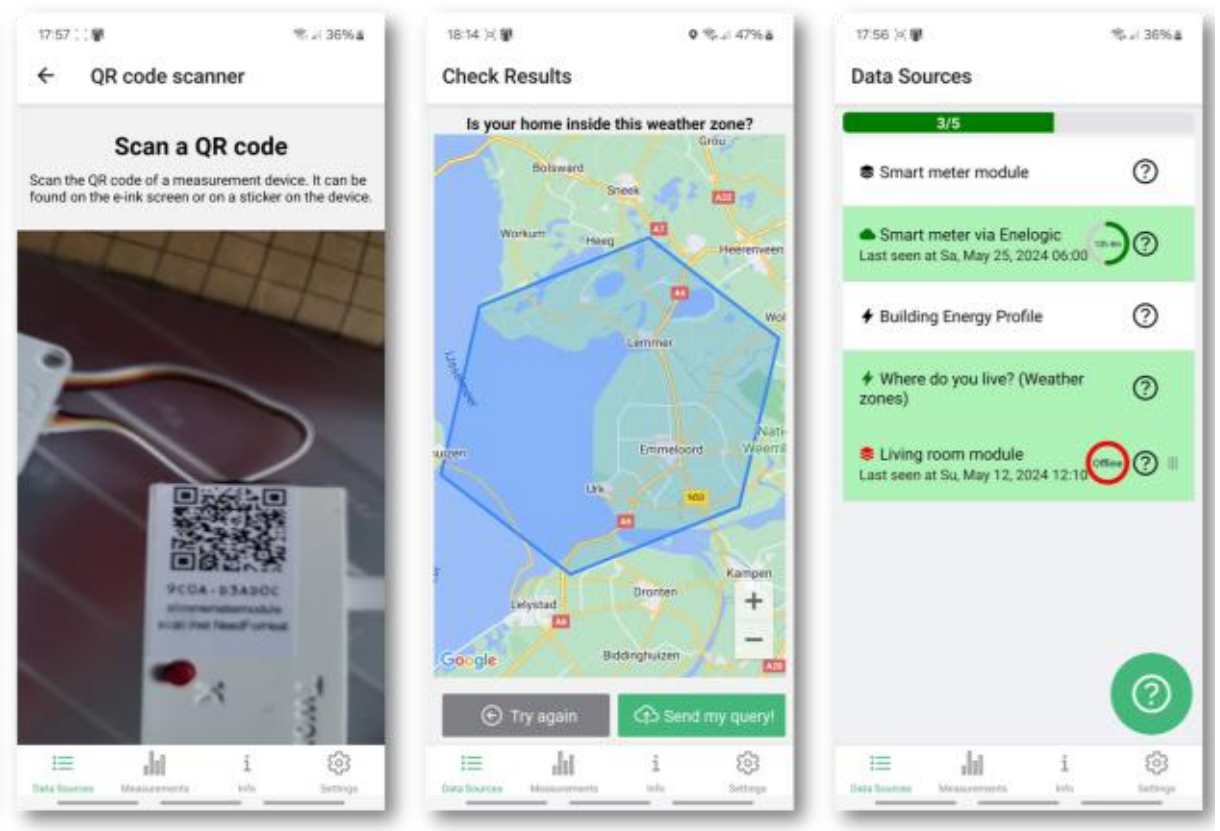

*Fig. 2. NeedForHeat GearUp App Screenshots*

With just two clicks, residents can download and install the app using a personal installation and activation link received via email. The embedded activation token in the link automatically logs into the NeedForHeat Server, retrieving campaign details and required data sources for display in the app (see Fig. 1). As this relies on pseudonyms, directly identifying information like email address is never stored on the NeedForHeat Server, in line with the privacy-by-design requirement mentioned in section II.C (see also section V.C).

The app currently supports the activation of three types of data sources, with a fourth type, batch import data, handled outside the app:

#### 1) NeedForHeat Measurement Devices

The app helps residents install and connect Measurement Devices received by mail. After scanning a device QR code, the app provides step-by-step installation guidance, including connection to home Wi-Fi. Once the first device is configured, network settings can be reused for additional devices.

#### 2) NeedForHeat Cloud Feeds

When campaigns require monitoring data from online data sources, such as Enelogic (an ODA that can retrieve smart meter data via P3/P4 without additional hardware), the app lets residents securely authorize NeedForHeat to access such Cloud Feeds, without sharing their credentials.

#### 3) NeedForHeat Energy Queries

Energy Queries are one-time information requests to obtain campaign data. Currently, two queries are implemented, both adhering to privacy-by-design principles:

- *Weather Zone Query*: Residents confirm their home location in the app. To protect privacy we use the H3 hierarchical hexagonal geospatial indexing system [20] with Gaussian noise to assign the home to an H3 level 4 cell (see Fig. 3). Only the H3 cell ID and time zone (e.g. 'Europe/Amsterdam') are sent to the Server. The server uses the H3 cell centre for geospatial interpolation of KNMI weather data [21] and the time zone for local time reconstruction.

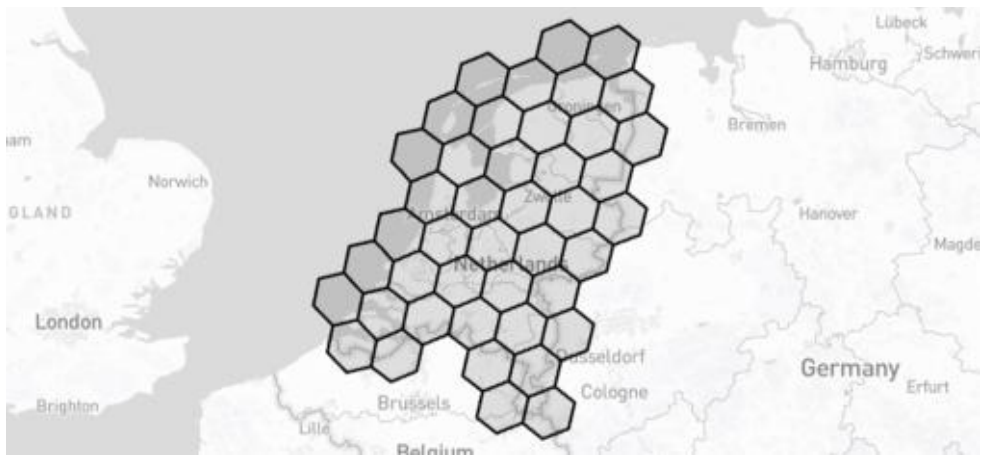

*Fig. 3. H3 level 4 hexagonal cells used as Weather Zones*

- *Building Performance Signature Query*: Residents confirm their home address, which is used by the app to retrieve data about their dwelling via open BAG API [22] and 3D-BAG API [23]. Privacy is enhanced by processing data in the app locally, before sending the results, a (3D-)BAG-data-derived building performance signature, to the NeedForHeat Server.

#### 4) *Batch Import Data*

This includes data, like KNMI data and Remeha boiler data, added later to the NeedForHeat Server database during diagnosis, outside the scope of GearUp.

As shown in Fig. 2 and in a video [24], GearUp also allows residents to monitor data collection health, showing the timestamp of the most recent measurement and time to the next measurement as well as visualisations and notifications of overdue data sources, ensuring smooth data collection.

### B. *NeedForHeat Measurement Devices*

NeedForHeat Measurement Devices collect heating and environmental data from homes, including energy, boiler, indoor conditions and occupancy data (see Appendix 2 for details).

#### 1) *NeedForHeat Boiler Module*

NeedForHeat Boiler Modules monitor gas-fired boilers by collecting data on operation and system temperatures, including water flow, return, room, and setpoint temperatures, to assess heating system efficiency and heat distribution characteristics before transitioning to new heating solutions. Measuring flow and return temperatures inside boilers (e.g., via OpenTherm Monitor or Remeha data) may be affected by hot water production. Pipe clamp measurements, less prone to such interference, can serve as a remedy.

*a) OpenTherm Monitor.* The OpenTherm Monitor, a compact device installed near the boiler, captures thermostat-boiler communication. GearUp guides residents through installation, cutting and stripping thermostat wires and connecting them to pluggable screw terminals. During de-installation, residents reconnect wires using a terminal block, restoring the original setup before returning the device.

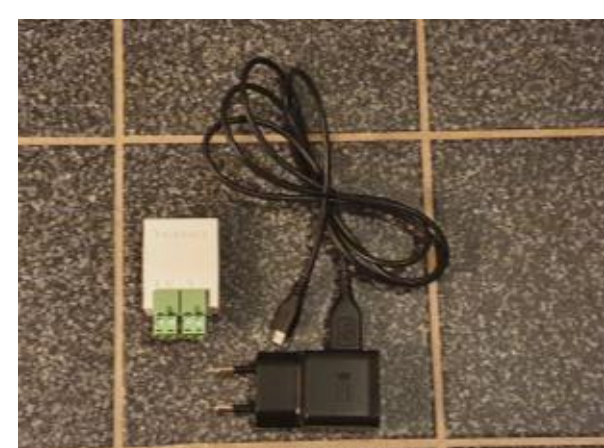
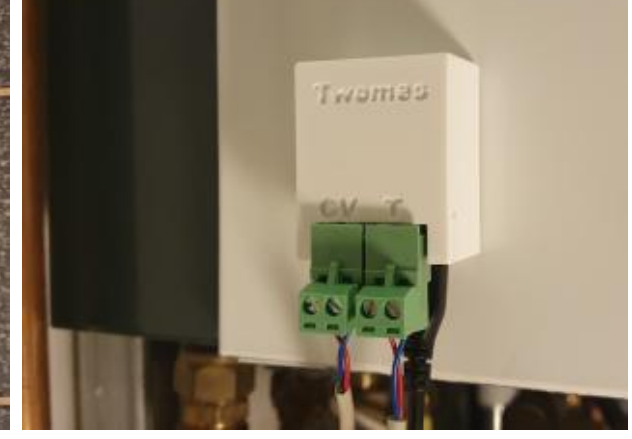

*Fig. 6. OpenTherm Monitor (left: shipped; right: installed)*

*b) Boiler Monitor Satellite.* The Boiler Monitor Satellite measures water flow and return temperatures to the heat distribution system (e.g. radiators), providing insights into heating efficiency and heat distribution system characteristics. This battery-operated device is easily installed by magnetically attaching it to the boiler and securing temperature sensors to pipes with clips. Data is sent wirelessly via ESP-NOW *[25]* to the Smart Meter Module, which forwards it to the NeedForHeat Server .

Based on experiences in field study A (see section VII.A), we started development of an integrated OpenTherm Monitor and Boiler Monitor Satellite (see section VIII.B), as a BASE module for the M5Stack CoreInk, which uses direct Wi-Fi transmission to the NeedForHeat Server.

#### 2) *Living Room Module*

The NeedForHeat Living Room Module measures indoor environmental conditions, including temperature, humidity, $CO_2$ concentration, and occupancy counts, typically at 10-minute intervals. As $CO_2$ levels rise with occupancy and decay through ventilation and infiltration, they provide a basis for distinguishing heat losses from ventilation, infiltration, conduction, and convection, particularly when combined with wind speed data. The module should be placed by the resident in the main living area, typically near the thermostat, and should be connected to mains power.

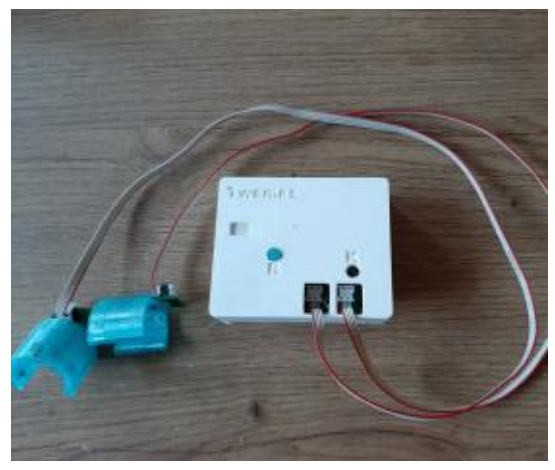
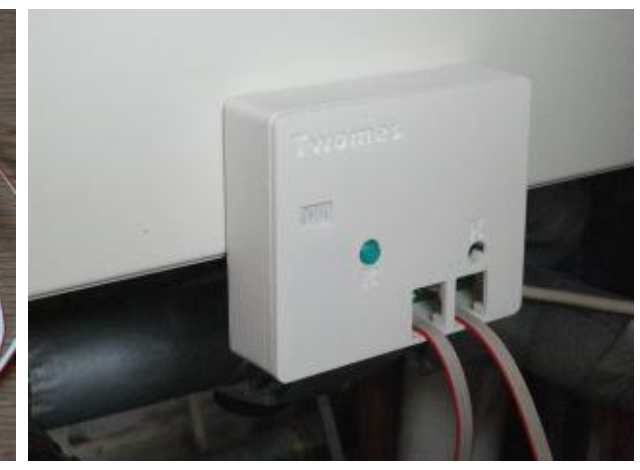

*Fig. 4. Boiler Monitor Satellite (L: shipped; R: installed)*

The module also performs occupancy detection by sending Bluetooth name requests and counting registered smartphones in range while ignoring others. At the start of a measurement campaign, each household member registers their smartphone once; the module guides them through a one-time pairing process. After setup, the module operates autonomously, requiring no further action beyond keeping Bluetooth enabled. Given that over 95% of Dutch residents aged 12+ own a smartphone [2], this method ensures broad applicability. The module builds on experience in field study A with the Room Temperature Satellite [26]-[27], which collected similar data, except for occupancy, gathered by the OpenTherm Monitor.

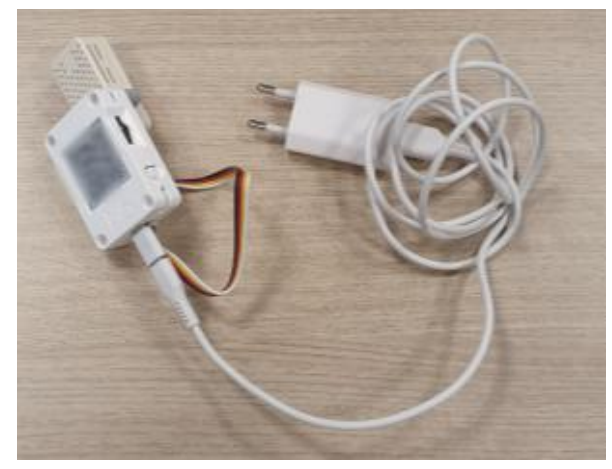
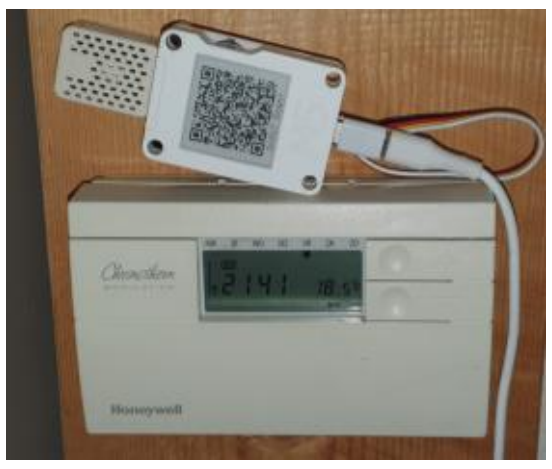

*Fig. 5. Living Room Module (left: shipped; right: installed)*

#### 3) *Smart Meter Module*

The NeedForHeat Smart Meter Module connects to the P1 port of Dutch smart meters, enabling the collection of electricity and gas meter readings from nearly all of the over 95 different smart meter types currently deployed in the Netherlands [28], typically at 10-minute intervals. This enables analysis of how efficiently energy flows are converted into heat, for example via central heating, or as residual heat from hot tap water or as a result of indoor electricity consumption. The module supports easy installation by residents, requiring only a smart meter connection, mains power for older DSMR3.0 meters, and configuration via the GearUp app. By default, cabling, a 230V mains power to micro-USB 5V power adapter, and an external P1 splitter are included, allowing continued use of existing P1 readers.

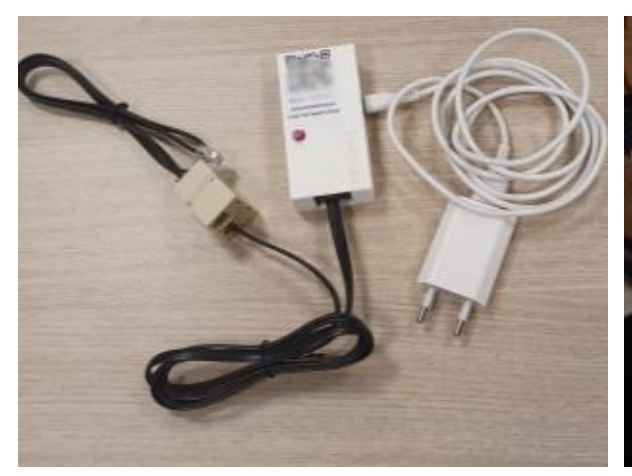
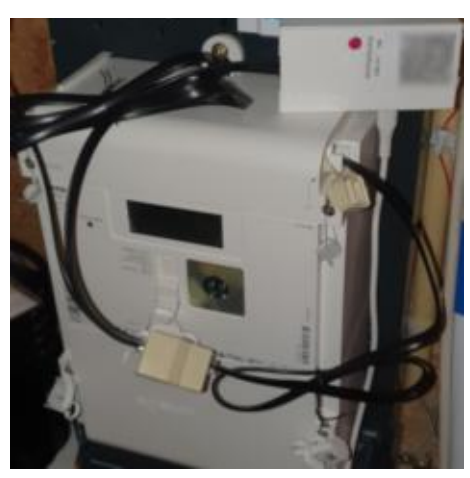

*Fig. 7. Smart Meter Module (left: shipped; right: installed)*

Following experiences in field study A, we started developing an improved version of this module (see section VIII.A), as a BASE module for the M5Stack CoreInk.

## V. Deployment Support

In this section, we describe the support that NeedForHeat DataGear provides for deployers that want to configure, deploy and monitor measurement campaigns, but that do not want to engage in extending or modifying NeedForHeat DataGear in a way that may involve programming, compiling, hardware (re)design, as described in section VI. The sections below describe the support NeedForHeat DataGear provides for each deployment step.

### A. Deploying a NeedForHeat Server Instance

The NeedForHeat Server is a cloud-agnostic modular platform deployed via Docker containers using open-source technologies for data collection and management. Unlike proprietary cloud solutions, it runs on any Infrastructure-as-a-Service (IaaS) or self-hosted environment. It requires a Linux server, a manageable DNS domain, and the NeedForHeat Server deployment manual, plus experience to follow instructions. Key containers include:

- **Portainer**: Server config and container management.
- **Traefik**: Secure access and reverse proxy.
- **NeedForHeat manual server**: nginx web server for installation manuals, FAQs, and privacy policies.
- **NeedForHeat Server API**: Dedicated REST API that facilitates provisioning and data uploads from measurement devices, supports measurement device provisioning and monitoring via the GearUp app.
- **MariaDB**: Database that stores measurement campaign configurations and measurement data, and secure access for NeedForHeat Diagnosis containers.
- **CloudBeaver**: Web-based database access.
- **Duplicati**: Incremental backups of measurement data to a storage site (e.g., SURFdrive, Research Drive).
- **JupyterLab**: NeedForHeat Diagnosis containers.

### B. Preparing Measurement Devices

Preparing measurement devices for deployment involves selecting and producing hardware components, as well as installing firmware. The NeedForHeat project offers open hardware designs that simplify this process, making it accessible even to deployers with limited hardware expertise.

#### 1) Producing Hardware Components

The NeedForHeat devices are available as open hardware designs, with fabrication files for both PCBs and 3D-printed enclosures. This allows for straightforward production of:

- **Printed Circuit Boards (PCBs)**: Fabrication files can be sent to various PCB manufacturing services, which will fabricate and assemble the boards as specified. This process is akin to submitting a print job and awaiting delivery in a few days.
- **3D-Printed Enclosures**: Fabrication files are available and can be sent to 3D printers or 3D printing services, which will produce the enclosures based on the provided specifications.

This methodology streamlines PCB and enclosure production, delivering ready-to-assemble components (see Fig. 8). Final steps include placing PCBs in enclosures and connecting cables before shipping to residents.

#### 2) Installing Measurement Firmware

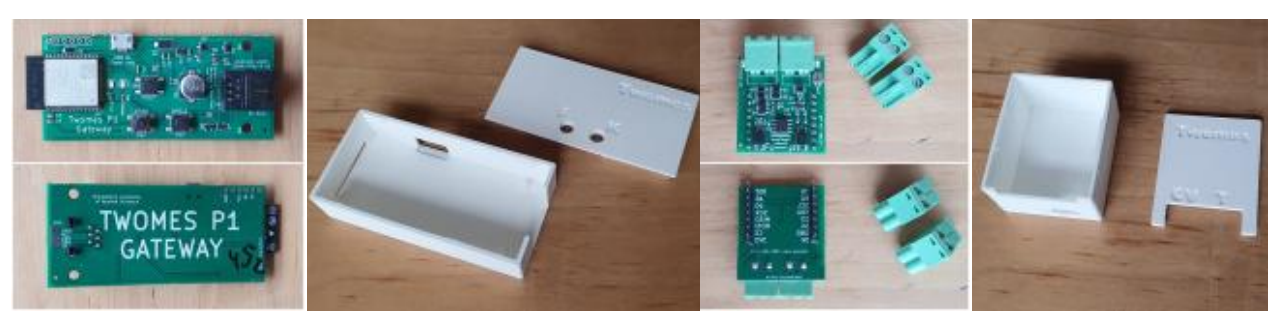

*Fig. 8. Smart Examples of PCBs and enclosures as produced*

Once the hardware components have arrived and have been assembled, the next steps are:

- **Firmware Installation**: The firmware tailored for each device type should be flashed onto the hardware, by connecting the device to a computer and using the tools provided.
- **Provisioning QR Codes**: Measurement devices without M5Stack CoreInk require printed QR codes for unique identification and provisioning. In contrast, devices that use the M5Stack CoreInk display the QR code on their e-ink screen, streamlining the process by eliminating the need for the laborious task of manual label printing and attachment.

#### 3) Customizing User Manuals and FAQs

Manuals and FAQs can be overridden:

- **Default Manuals**: The standard manuals and FAQs are available in English and Dutch in the firmware repositories.
- **Custom Manuals**: Deployers can provide campaign-specific user manuals by placing adapted versions in the desired language in the appropriate directory of the manual server. This customization ensures that residents receive information tailored to their specific deployment context .

#### 4) Overview of Developed Devices and Cost

In TABLE I. we provide an overview of devices we developed so far, including details on hardware production costs and costs associated with other gear, such as cables and power adapters. For a list of all data properties that can be collected, see Appendix 2.

TABLE I. Cost* of NeedForHeat Measurement Devices

| NeedForHeat Device | Cost per item | | | | Total |
|---|---|---|---|---|---|
| | ***PCB*** | ***Enclosure*** | ***CoreInk*** | ***other*** | |
| Living Room Module + $CO_2$ | - | - | € 36 | € 52 | € 88 |
| Living Room Module - $CO_2$ | - | - | € 36 | € 15 | € 51 |
| Smart Meter Module | € 10 | € 5 | - | € 17 | € 32 |
| OpenTherm Monitor | € 4 | € 2 | - | € 13 | € 19 |
| PostNL | - | - | - | - | € 4 |
| P1-BASE | € 12 | € 6 | € 36 | € 5 | € 59 |
| Boiler Monitor | € 19 | € 11 | € 36 | € 23 | € 89 |

[a] *Approximate prices (2021-2025) excl. VAT; prices vary per supplier and order quantity*

### C. *App and Device Distribution*

NeedForHeat DataGear supports distributing the measurement app and devices to residential participants. The NeedForHeat GearUp app (see section IV.A) in the Google Play and Apple App Stores, works with custom measurement campaigns hosted on any NeedForHeat Server instance.

To onboard residents, deployers manually create campaigns and accounts via the NeedForHeat Server API (see Section V.A), using web-based interactive API calls using the API's Swagger GUI, specifically POST on `/app`, `/campaign`, and `/account`. These calls return an account activation token, which must be URL-encoded into a personal app installation and activation link for each resident. Since the activation token is deleted after use, the researcher creating the accounts must generate a pseudonym for each resident and link it to the account id. The helpdesk and researcher must coordinate to ensure each account id, activation token, and pseudonym are correctly assigned and recorded in the pseudonym-to-resident key file that is securely stored outside of the NeedForHeat Server. Future enhancements, such as the NeedForHeat Cockpit web interface (see section IX), aim to simplify these processes.

The helpdesk, which maintains the pseudonym-to-resident key file, is responsible for securely emailing each resident their personal activation link and shipping measurement devices to their respective street addresses.

### D. *Monitoring Data Collection*

Currently, NeedForHeat DataGear provides little to no automated support for monitoring whether they answered the Energy Queries, devices were installed and activated correctly, and the devices are uploading data at the expected intervals.

Deployers must manually retrieve this information via the CloudBeaver web interface, using data exports or custom SQL queries. Future enhancements, such as the NeedForHeat Cockpit (see section IX), aim to simplify and streamline this process.

## VI. Development Support

NeedForHeat DataGear supports developers in modifying or extending the firmware and hardware components. All NeedForHeat designs are available open source on GitHub [29] under permissive licenses: software under Apache 2.0 and hardware under CERN-OHL-P v2. Developers are encouraged to contribute improvements via Pull Requests or new repositories with similar licenses. To accommodate commercial parties in the energy transition, the licenses allow commercial use and proprietary modifications without a legal obligation to share changes.

### A. *NeedForHeat Generic Device Firmware*

NeedForHeat Generic Device Firmware [30] supports IoT devices with ESP32, leveraging Bluetooth and Wi-Fi. As shown in Fig. 9, it builds on ESP-IDF, adding features like Provisioning, Time Syncing, Heartbeats, Secure Upload, Occupancy Counting, Persistent Buffering, and OTA firmware updates.

#### 1) *NeedForHeat Provisioning*

Built on Espressif Unified Provisioning [31], NeedForHeat Provisioning uses BLE for Wi-Fi setup. The device QR-code's name field uses a hash of device type name and MAC address, allowing the GearUp app to retrieve the installation manual. The Proof-of-Possession field in the QR-code derives from the ESP32 hardware RNG to prevent MITM attacks. On the M5Stack CoreInk, the QR-code displays on the e-ink screen. After provisioning, the device calls `/device/activate`, preventing duplicate activation. Once the first heartbeat uploads, the GearUp app signals successful provisioning.

#### 2) *Scheduling & Heartbeats*

The firmware schedules heartbeats, measurements, and uploads while optimizing power use. The ESP32 switches off or enters light sleep based on a calculated threshold. If developers provide energy and duration metrics (e.g., boot/setup power and time, sleep/off mode power draw), the firmware computes the most efficient power-saving strategy. On the M5Stack CoreInk, only the BM8563 RTC remains active in off mode, drawing 2.3 µA, resulting in a power-off threshold of 1-2 minutes before the next scheduled task.

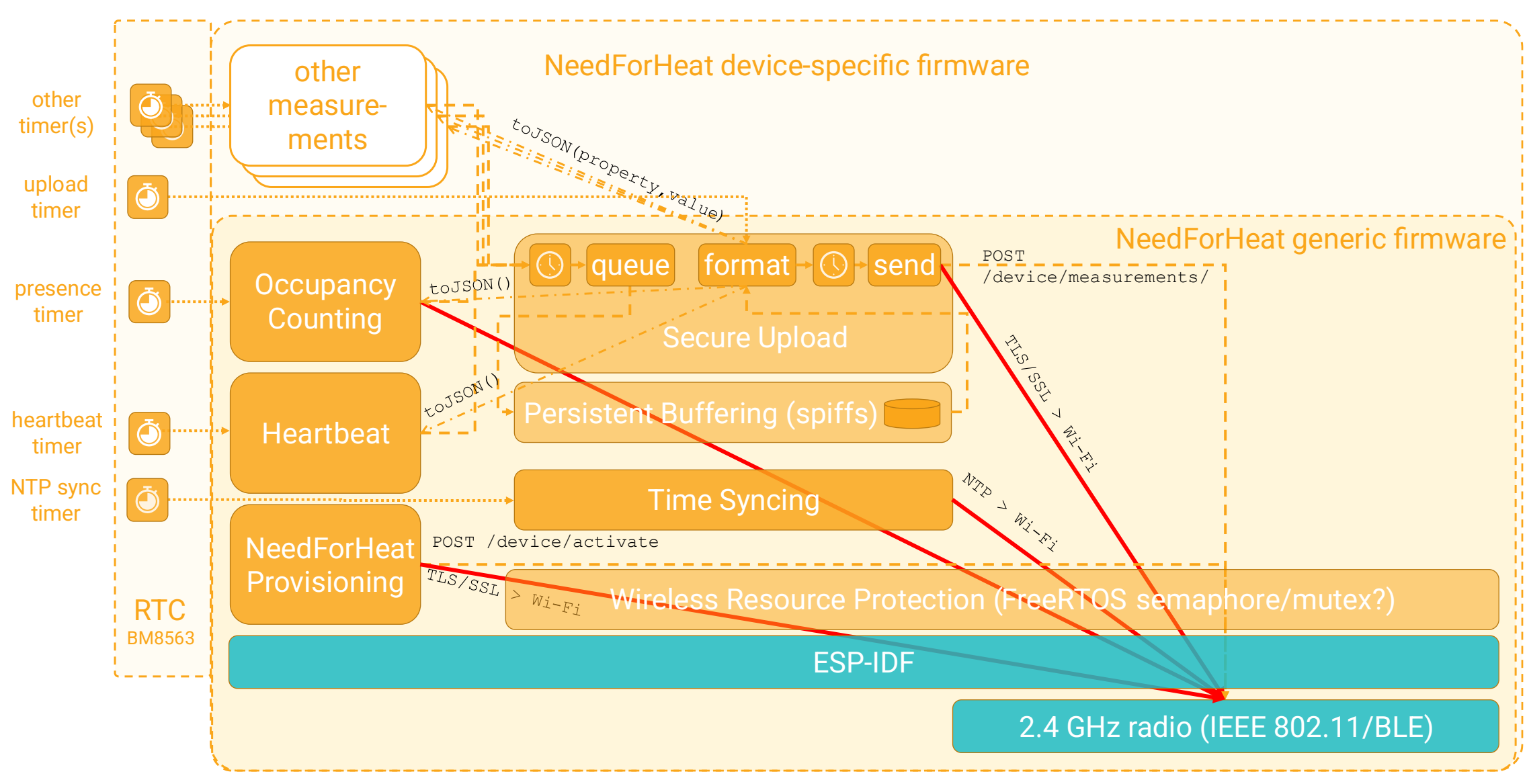

*Fig. 9. Generic Device Firmware, process view*

*3) Time Synchronization*

Shortly after provisioning, a measurement device syncs its device clock with an NTP server. This repeats daily. On M5Stack CoreInk, time can also be read from the BM8563 RTC.

*4) Secure Upload*

Measurements are timestamped in Unix time (the Smart Meter Module converts P1 timestamps to Unix time, handling DST transitions, even on DSMR3.0 devices during ambiguous hours), stored in a buffer, formatted as JSON, and uploaded with an upload timestamp to the NeedForHeat Server.

*5) Occupancy Counting*

Inspired by monitor [32], occupancy counting firmware sends Bluetooth name requests to registered MAC addresses, typically at 10-minute intervals. On the M5Stack CoreInk, a menu guides residents to register by pairing their phone with '`NeedForHeat_OK`'. Registered phones with Bluetooth enabled are counted: they respond automatically to requests.

*6) Persistent Buffering*

The firmware supports storing measurements (typically at 10-minute intervals) and bulk-uploading (e.g., every 6 hours), enabling power savings and robustness against short internet outages. This feature has not yet been field-tested.

*7) Secure Over-The-Air (OTA) firmware updates*

OTA firmware updates were implemented but not tested.

### B. NeedForHeat Server Database and API

The NeedForHeat database model supports data collection and campaign management (see Appendix 1 for the ER diagram). An `Account` represents a resident participating in a data collection `Campaign` via an `App`, which facilitates activation of items in the `DataSourceList`, which specifies the types of data being gathered in the `Campaign`. Each `DataSource` is linked to a `DeviceType`, an `EnergyQuery` or a `CloudFeed`. These `DataSources` generate `Measurements`, which are bundled in `Uploads`. Each `Measurement` records a timestamped value for a physical or environmental `Property`, with a well-defined convention how units are encoded in `Property` names [33].

The NeedForHeat Server API follows a RESTful architecture and provides secure endpoints for:

- **Authorization**: manage admin, app & device access.
- **Campaigns**: manage apps, campaigns & data sources.
- **Provisioning**: account, device & Cloud Feed activation.
- **Uploads:** measurements and Energy Query results.
- **Retrieval**: querying device status.

### C. NeedForHeat GearUp app

The app's open-source code allows developers to customize branding, languages, and energy queries as needed.

## VII. Case Studies

NeedForHeat DataGear has been used in three field studies so far, testing the system in different contexts and gaining insights in data reliability, usability, and robustness (see TABLE II. ) .

TABLE II. NeedForHeat Deployment Overview

| *Field Study* | *Context* | *Year* | *#wks* | *#hh sent* | *#hh valid* | *#data points* |
|---|---|---|---|---|---|---|
| A | Residential | '21-'22 | 21 | 40 | 18 | 35,0M |
| B | Office | '22-'23 | 3 | 6 | 3 | 386k |
| C | Residential | '23-'25 | 21 | 42 | 20 | 30,9M |

### A. Field Study A: Residential Study

Field Study A targeted pre-transition homes without smart thermostats. We observed a large variety of heating systems: 35 unique thermostat-boiler combinations across 40 homes. Support for OpenTherm return temperature varied, crucial for boiler efficiency analysis. Most residents self-installed the devices, but some required assistance from local volunteers or researchers. Key lessons and improvements:

- *Simplified provisioning:* Printing device-specific QR-codes proved cumbersome, and 3D-printed enclosures with loose buttons were vulnerable during shipping. Two categories of devices (primary and satellite) complicated device installation. We switched to PostNL letterbox parcels for shipping, and M5Stack CoreInk hardware featuring an e-ink screen for QR codes, integrated buttons and rechargeable battery. Additional hardware is now designed as a CoreInk BASE module. The new server API v2.0 allows any device to be sent to any home.
- *Integrated Boiler Monitor:* Intake surveys required detailed questions to determine OpenTherm compatibility. To streamline deployment, we started designing a single device for OpenTherm and other heating systems.

### B. Field Study B: Office Spaces

A short-term office study explored $CO_2$ levels, occupancy, and ventilation losses in offices [34, pp. 53–78]. While small in scale, results were encouraging for ventilation-related heat loss modelling. Despite optimized scheduling, the 390 mAh CoreInk battery lasted only 5 days. We began designing a 3xAAA BatteryBASE, but ultimately switched to mains power for Living Room Modules in later studies.

### C. Field Study C: Residential Study with Smart Thermostats

Unlike Field Study A, all homes in Field Study C had Remeha boilers and smart thermostats, eliminating the need for boiler monitors, or so we thought. Domestic hot water usage interfered with flow and return temperature readings, requiring a complicated filtering algorithm for preprocessing. External pipe clip sensors could have helped validate this approach.

Additionally, 1-minute interval thermostat and boiler data relied on a diagnostic mode that frequently reset, creating data gaps. Remeha implemented a manual profile restoration check to minimize losses. An automated solution has since been developed, but the dataset still contains many gaps.

## VIII. Devices Under Development

Based on experiences from the field studies, we initiated a migration of all measurement devices to add-on BASE modules compatible with the M5Stack CoreInk. At the time of writing, two such modules were still under development. We welcome Pull Requests from collaborators interested in completing these designs.

### A. P1-BASE Module

An updated P1-BASE module is under development as a BASE module for the M5Stack CoreInk. It integrates a P1 splitter to reduce cable clutter and is designed for improved compatibility with Landis+Gyr E360 smart meters.

The current design [35] still includes DIP switches to configure the P1 data request line (always on, via the P1-BASE, and/or via an external smart meter reader). Future iterations aim to replace these hardware switches with software-controlled configuration to improve usability.

### B. Integrated Boiler Monitor

To simplify deployment and eliminate the need for pre-screening OpenTherm compatibility between boilers and thermostats, an Integrated Boiler Monitor is being developed. In earlier projects, researchers relied on intake surveys in which residents manually selected boiler and thermostat models and submitted photographs, an error-prone and time-consuming process. The new module is intended to function in any home, removing this step entirely.

The design [36] consists of a BoilerBASE module serving as a BASE module for the M5Stack CoreInk. The BASE module connects to pipe-mounted temperature sensors via RJ-12 cables, optionally using thermally conductive pads to improve thermal contact. The BASE module also interfaces with a thermostat cable splitter via an RJ-45 connection. The splitter features two 4-position pluggable screw terminals allowing residents to connect two to four thermostat wires in any order. For de-installation, residents simply unplug the RJ-45 connector from the thermostat cable splitter, leaving the splitter in place and returning the remaining components.

For homes with wireless thermostats, future versions may include an 868 MHz transceiver to monitor wireless communication between boiler and thermostat. As this requires explicit user consent, pairing would be guided via the GearUp app.

## IX. Contributions and Limitations

NeedForHeat DataGear enhances residential energy monitoring by bridging gaps in existing solutions. Unlike smart thermostats (e.g., Toon [13], Honeywell [14], Nest [15]), which provide only indoor temperatures and setpoints, NeedForHeat captures a broader range of heating-related parameters. Unlike platforms like BeNext [37], which require a special gateway, NeedForHeat only requires Wi-Fi, reducing deployment barriers.

Compared to electricity-focused monitoring systems like HomeWizard [38], HanzeBox [39], OpenEnergyMonitor [40], and IUNGO [41], NeedForHeat also collects gas usage, flow and return temperatures, and thermostat interactions. Unlike home automation platforms like Home Assistant [19] and Homey [42], it prioritizes scientific-grade data collection over real-time control, offering reliable and accurate timestamps useful for modelling building thermal dynamics rather than solely providing energy feedback.

Compared to monitoring systems, like BeNext [37], and Watch-E [43], which focus on post-installation compliance for norms like EPV and NOM-keur, NeedForHeat supports pre-transition diagnostics, enabling tailored advice.

The resources page of ACM SIGEnergy [44] lists a broad collection of modelling and simulation tools as well as datasets relevant to energy systems and energy informatics research. While it includes software toolkits and data for residential energy studies, it does not catalogue standalone open hardware platforms or dedicated open-source software suites for residential energy data collection. NeedForHeat DataGear addresses this gap by offering an open-source platform for residential energy monitoring.

Despite these contributions, NeedForHeat lacks standardized interoperability with the Energy Performance Monitoring API [45], [46] or semantic data schemas like Brick [47] and Haystack [48]. Deployment remains challenging. Currently, the process requires manual account creation, pseudonym assignment, and installation tracking; tasks that NeedForHeat Cockpit, now in early design, will support. Real-time campaign health monitoring needs further development to identify installation and data transmission issues.

## X. Conclusions

NeedForHeat DataGear provides an open, flexible, and scalable data collection platform for residential heating research, addressing limitations in existing systems. Its scientific-grade timestamping, modular design, and broad heating parameter coverage enable pre-transition diagnostics, and tailored advice. Early deployments demonstrated data reliability and usability, but challenges remain in standardization, automation, and deployment support. As open software, NeedForHeat could support future enhancements in interoperability, onboarding, and real-time monitoring, enabling broader contributions to large-scale studies.

## Acknowledgments

This work was supported by Centre of Expertise TechForFuture, BDR Thermea Group, Enpuls, Breman Installatiegroep, Witteveen+Bos, and Gemeente Zwolle. We thank Arjan Peddemors from Aquilis for contributions to the initial design and implementation of the NeedForHeat server infrastructure and API. Special thanks to Marco Winkelman from Windesheim, for his contributions to hardware and electronics design beyond regular student guidance. We also acknowledge the students from Windesheim who contributed to NeedForHeat DataGear, as part of a learning assignment: Meriyem Bilgili, Laurens de Boer, Jesse Brand, Thomas Buijs, Gerwin Buma, Bram Busch, Huub Buter, Jort Driegen, Niels Even, Kees Fokker, Rick de Graaf, Gerwin ten Have, Werner Heetebrij, Brian Hoen, Kevin Jansen, Tristan Jansen, Tristan Jongedijk, Leon Kampstra, Jorrin Kievit, Rick Klaasboer, Nick Koster, Erik Krooneman, Joël Kuijper, Fredrik-Otto Lautenbag, Robin Leuninge, Thomas van Meer, Harris Mesic, Tiemen Molenaar, Matthijs Noordhof, Ward Pieters, Marco Prins, Amicia Smit, Sjors Smit, Daan Tellegen, Wietske Veneberg, Maarten Vermeulen, Matthias Verweij, Mirjam van Wee, Joël van de Weg, Stijn Wingens, and Rowan van der Zande. Finally, we thank xAI's Grok and OpenAI's ChatGPT for assistance with English formulations and sentence clarification.

# APPENDIX 1. NEEDFORHEAT DATABASE ER DIAGRAM

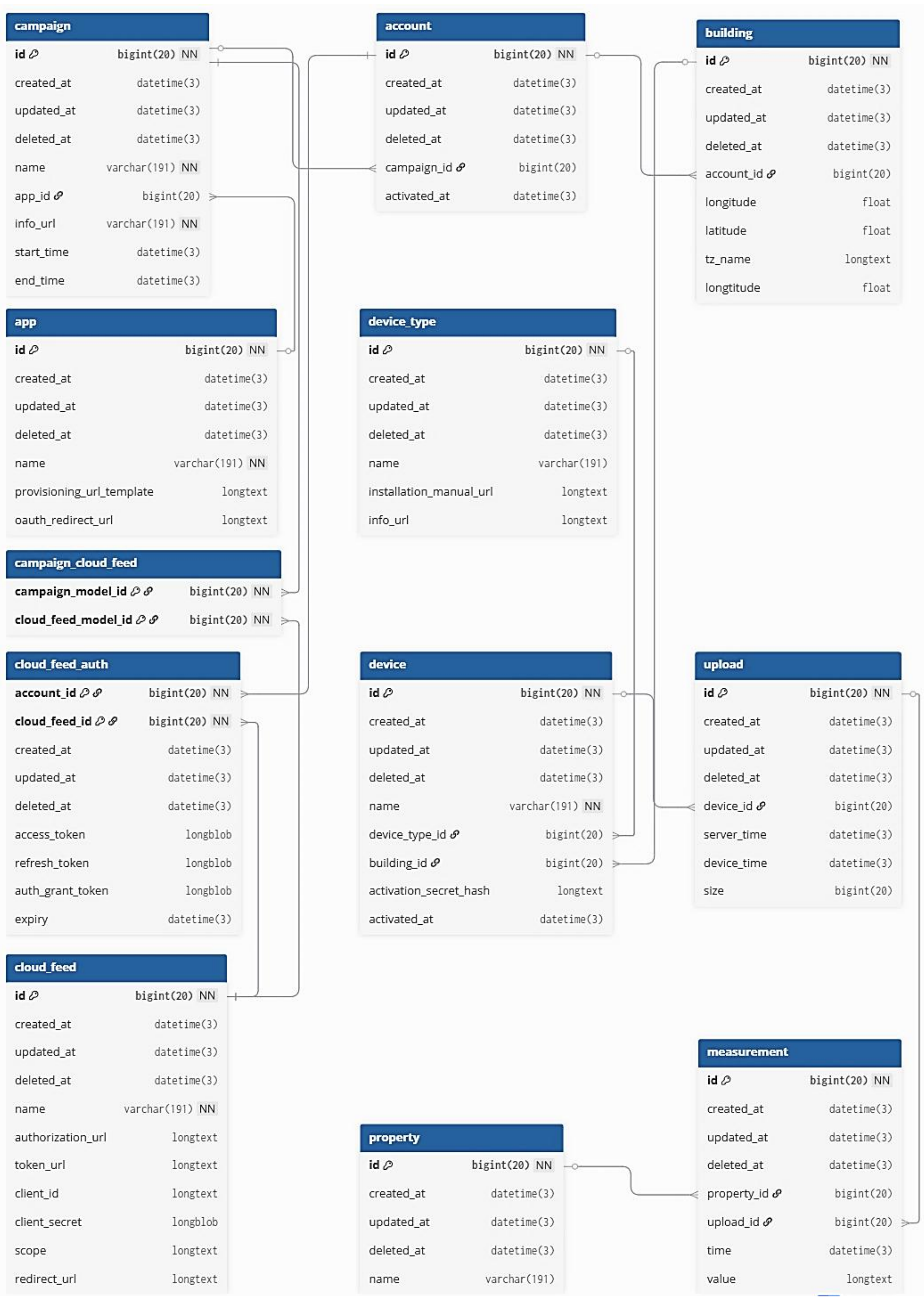

## Appendix 2. NeedForHeat Devices and Properties

| source name | interval | property | description | unit | format | sensor |
|---|---|---|---|---|---|---|
| **Living Room Module** | 0:10:00 | heartbeat__0 | heartbeat counter | [-] | %u | ESP32 |
| | 0:10:00 | co2__ppm | indoor $CO_2$ concentration | ppm | %u | Sensirion/SCD41 |
| | 0:10:00 | temp_indoor__degC | indoor air temperature | ˚C | %.1f | Sensirion/SCD41 |
| | 0:10:00 | rel_humidity__0 | indoor relative humidity | [-] | %.1f | Sensirion/SCD41 |
| | 0:10:00 | onboarded__p | #onboarded smartphones | persons | %u | ESP32/Bluetooth |
| | 0:10:00 | occupancy__p | #onboarded smartphones seen | persons | %u | ESP32/Bluetooth |
| **Smart Meter Module** | 0:10:00 | heartbeat__0 | heartbeat counter | - | %u | ESP32 |
| | 0:10:00 | e_use_hi_cum__kWh | electricity meter reading | kWh | %.3f | P1 port |
| | 0:10:00 | e_use_lo_cum__kWh | electricity meter reading | kWh | %.3f | P1 port |
| | 0:10:00 | e_ret_hi_cum__kWh | electricity meter reading | kWh | %.3f | P1 port |
| | 0:10:00 | e_ret_lo_cum__kWh | electricity meter reading | kWh | %.3f | P1 port |
| | 0:10:00[1] | g_use_cum__m3 | gas meter reading | m3 | %.3f | P1 port |
| | 0:10:00 | meter_code_str | meter type code | n/a | %s | P1 port |
| | 0:10:00 | dsmr_version__0 | DSMR version | [-] | %.1f | P1 port |
| **OpenTherm Monitor**[2] | 00:00:30 | isBoilerFlameOn | status /flame status | bool | %d | OpenTherm |
| | 00:00:30 | isCentralHeatingModeOn | status/ch mode | bool | %d | OpenTherm |
| | 00:00:30 | isDomesticHotWaterModeOn | status/dhw mode | bool | %d | OpenTherm |
| | 00:00:30 | maxModulationLevel | capacity setting | % | %d | OpenTherm |
| | 00:00:30 | maxBoilerCap | max capacity | kW | %d | OpenTherm |
| | 00:00:30 | minModulationLevel | min-mod-level | % | %d | OpenTherm |
| | 00:00:30 | relativeModulationLevel | relative modulation level | % | %d | OpenTherm |
| | 00:00:10 | boilerSupplyTemp | boiler water temp. | °C | %.2f | OpenTherm |
| | 00:00:10 | boilerReturnTemp | return water temperature | °C | %.2f | OpenTherm |
| | 00:05:00 | roomSetpointTemp | room setpoint | °C | %.2f | OpenTherm |
| | 00:05:00 | roomTemp | room temperature | °C | %.2f | OpenTherm |
| | 00:05:00 | boilerMaxSupplyTemp | max ch water setpoint | °C | %.2f | OpenTherm |
| **Boiler Monitor Satellite** | 00:00:10 | boilerTemp1 | flow/return pipe temperature | ˚C | %.1f | DS18b20 |
| | 00:00:10 | boilerTemp2 | flow/return pipe temperature | ˚C | %.1f | DS18b20 |
| **Room Monitor Satellite** | 00:05:00 | CO2concentration | indoor $CO_2$ concentration | ppm | %d | Sensirion/SCD41 |
| | 00:05:00 | roomTempCO2 | indoor air temperature | ˚C | %.1f | Sensirion/SCD41 |
| | 00:05:00 | humidity | indoor relative humidity | % | %.1f | Sensirion/SCD41 |
| | 00:05:00 | roomTemp | indoor air temperature | °C | %.1f | Si7051 |

[1] Smart meters with DSMR ≥ 5.0 register meter readings every 5 minutes, smart meters with DSMR < 5.0 register gas meter readings only every hour.

[2] Properties of the OpenTherm Monitor, Boiler Monitor Satellite, Room Monitor Satellite do not adhere to the convention how units are encoded in Property names [33].